\documentclass[article,twocolumn,superscriptaddress,amsmath,amssymb,prb]{revtex4-1}
\usepackage{amsmath, amssymb}
\usepackage[colorlinks,allcolors=blue]{hyperref}
\usepackage{graphicx}
\usepackage{color}
\newcommand{\op}[1]{\boldsymbol #1}

\setcitestyle{super}

\definecolor{burntorange}{rgb}{0.8, 0.33, 0.0}

\begin{document}
\title{\sffamily Quantum control of an oscillator using stimulated nonlinearity}

\author{Andrei Vrajitoarea}
\email{andreiv@princeton.edu}
\affiliation{Department of Electrical Engineering, Princeton University, Princeton, NJ 08540, USA}
\author{Ziwen Huang}
\affiliation{Department of Physics and Astronomy, Northwestern University, Evanston, IL 60208, USA}
\author{Peter Groszkowski}
\thanks{Present Address: Institute for Molecular Engineering, University of Chicago, Chicago, Illinois 60637, USA.}
\affiliation{Department of Physics and Astronomy, Northwestern University, Evanston, IL 60208, USA}
\author{Jens Koch}
\affiliation{Department of Physics and Astronomy, Northwestern University, Evanston, IL 60208, USA}
\author{Andrew A.\ Houck}
\email{aahouck@princeton.edu}
\affiliation{Department of Electrical Engineering, Princeton University, Princeton, NJ 08540, USA}

\date{Oct 23, 2018}

\maketitle 
\textbf{\sffamily
Superconducting circuits extensively rely on the Josephson junction as a nonlinear electronic element for manipulating quantum information and mediating photon interactions. Despite continuing efforts in designing anharmonic Josephson circuits with improved coherence times, the best photon lifetimes have been demonstrated in microwave cavities. Nevertheless, architectures based on quantum memories need a qubit element for addressing these harmonic modules at the cost of introducing additional loss channels and limiting process fidelities. This work focuses on tailoring the oscillator Hilbert space to enable a direct Rabi drive on individual energy levels. For this purpose we implement a flux-tunable inductive coupling between two linear resonators using a superconducting quantum interference device. We dynamically activate a three-wave mixing process through parametric flux modulation in order to selectively address the lowest eigenstates as an isolated two-level system. Measuring the Wigner function confirms we can prepare arbitrary states confined in the single photon manifold, with measured coherence times limited by the oscillator intrinsic quality factor. This architectural shift in engineering oscillators with stimulated nonlinearity can be exploited for designing long-lived quantum modules and offers flexibility in studying non-equilibrium physics with photons in a field-programmable simulator. 
}

Quantized electromagnetic excitations in superconducting circuits have become a promising platform for processing quantum information \citep{Devoret2013}. A central piece to this hardware is the Josephson effect, which grants nonlinearity with negligible dissipation. 
Nonlinear mesoscopic oscillators have been attracting much interest in studying non-equilibrium physics \citep{Dykman2012}. Electrical circuits composed of Josephson junctions behave as quantum systems with discrete energy levels resembling artificial atoms \citep{Clarke1988}. Although the coherence properties of anharmonic Josephson circuits have been improving these past two decades, storing microwave photons in harmonic oscillators leads to significantly longer coherence times \citep{Reagor2016, Ofek2016}. This promising development has launched efforts in building hardware-efficient architectures based on quantum memories \citep{Mirrahimi2014, Naik2017}. Since it is impossible to selectively address individual transitions in a harmonic system, Josephson qubits are used for preparing and transferring quantum information with negative impact on coherence and gate fidelities. In this work we present a new paradigm in exploiting the Josephson effect to perform logical operations directly on the oscillator Hilbert space using a dynamically activated nonlinearity.

\begin{figure*}[!t]
	\begin{center}
		\includegraphics[width=1.0\textwidth]{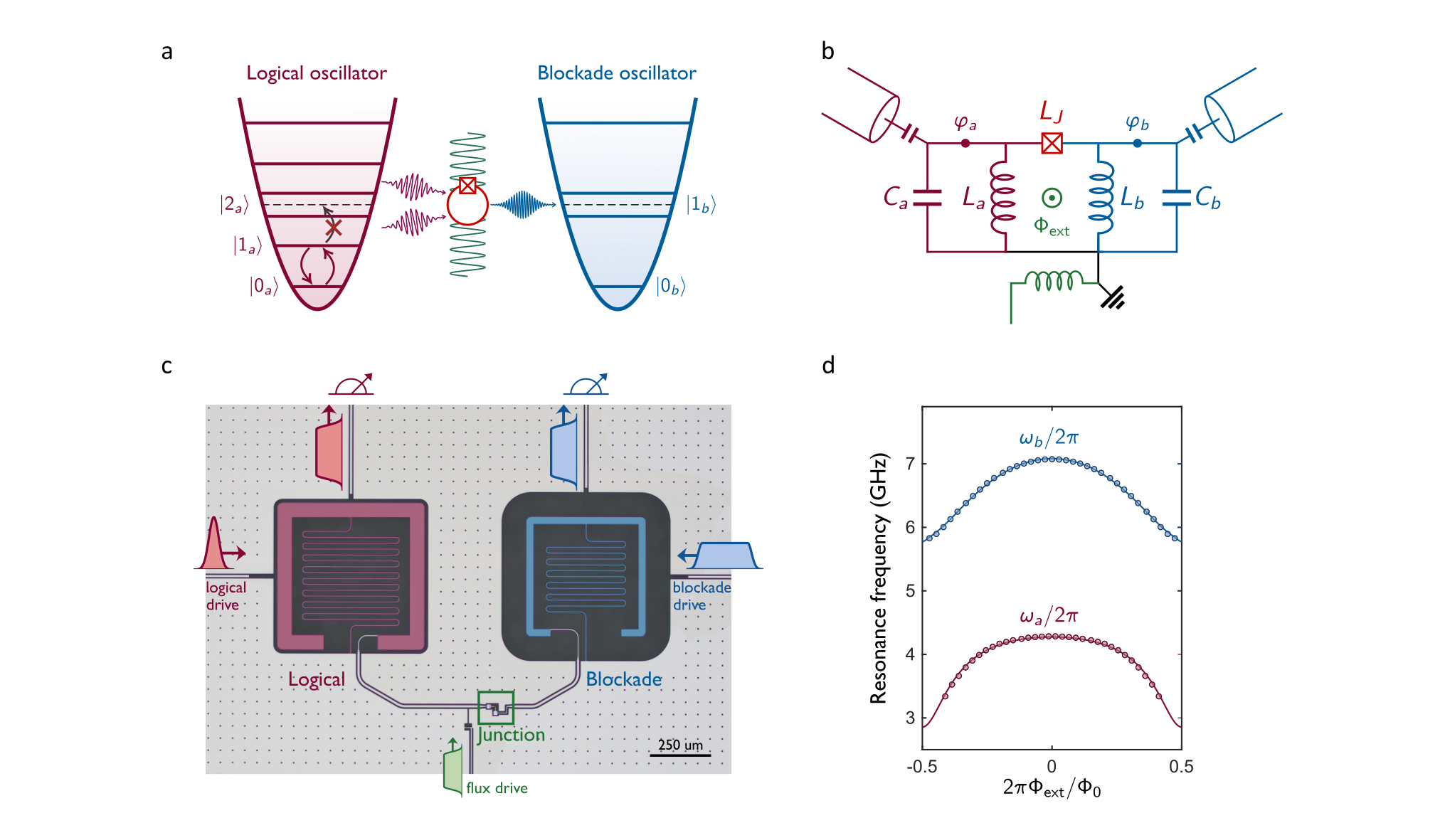}
	\end{center}
	\vspace{-0.6cm}
	\caption{\label{fig:Device} \textbf{Device description and spectroscopy.} \textbf{a.} Schematic energy spectrum for two coupled oscillators. Periodically modulating the coupling element stimulates a three-wave frequency conversion process which leads to a nonequidistant energy ladder. \textbf{b and c.} Schematic circuit diagram and false-color optical micrograph of the lumped element resonators, logical (red) and blockade (blue), coupled inductively through an rf-SQuID (green). Each cavity has a drive port and an output port for transmission measurements. The rf-SQuID has a broadband driving port used for sending the dc flux bias and the parametric pump tone. \textbf{d.} Measured resonance frequencies as a function of applied coupler dc flux bias. The circuit model (solid line) yields the fitted parameters: $C_a = 450.0$ fF, $C_b = 244.3$ fF, $L_a = 3.77$ nH, $L_b = 2.45$ nH, $L_J = 11$ nH.}
\end{figure*}

The key concept behind this experiment is to engineer a three-wave interaction between the electromagnetic modes of two linear oscillators, which we refer to as \textit{logical} and \textit{blockade} oscillator (Fig.~\ref{fig:Device}A). Parametrically modulating the interaction amplitude stimulates the conversion of two logical photons into one blockade photon and vice-versa. Provided the conversion rate is larger than the oscillator dissipation, this stimulated process strongly hybridizes the logical two-photon state with the blockade single-photon state, resulting in an anharmonic energy spectrum which allows selective control over the single-photon manifold of the logical mode. This conversion process has been previously demonstrated to implement a degenerate parametric oscillator used for stabilizing the oscillator state into a quantum manifold using the quartic nonlinearity of a transmon qubit \citep{Leghtas2015, Touzard2018}. Our experiment focuses on engineering a Josephson circuit with cubic nonlinearity which would yield a larger conversion rate, thus maximizing the anharmonicity.

In contrast to conventional qubit designs where the tunnel junction is used primarily to induce a non-parabolic inductive potential, this work exclusively employs the Josephson nonlinearity for perturbing higher-lying eigenstates while maintaining a parabolic potential to still benefit from the intrinsic oscillator coherence. Dynamically confining a harmonic oscillator to a subset of energy levels has been demonstrated in superconducting circuits using the back action of a driven ancilla qubit, known as quantum Zeno dynamics \citep{Bretheau2015}. However, the selectivity of the blocking tone is limited by the dispersive shift magnitude and parasitic leakage in the ancilla, making this approach unfeasible for implementing a qubit.

Our central tool for stimulating anharmonicity is parametric driving. This process is generally described by engineering a time-varying Hamiltonian for a coupled mode system with tunable degrees of freedom such as the coupling elements or mode resonance frequency, and periodically driving the system at the frequency detuning between energy states to be coupled. Parametrically activated interactions have been extensively employed in superconducting circuits, with applications in quantum computing, specifically in quantum gates \citep{Bertet2006, Niskanen2007, McKay2016} and parametric multi-mode architectures \citep{Naik2017, Reagor2018}, frequency conversion \citep{Bajjani2011}, dissipative stabilization \citep{Lu2017}, quantum limited amplification \citep{Bergeal2010, Castellanos2008, Yamamoto2008} and non-reciprocal signal processing \citep{Lecoq2017}. This experiment extends the current toolbox with the functionality of dynamically inducing nonlinearity with a purpose of engineering highly coherent qubits.

This dynamical toolbox can also be exploited for quantum simulation, where recent efforts have demonstrated artificial gauge fields for photons \citep{Roushan2017, Peropadre2013}. The Josephson effect can be harnessed as a resource for mediating interactions between microwave photons \citep{Bishop2009}. Our approach for parametrically controlling these interactions in a single oscillator can be extended to an array of coupled cavities, a promising testbed for investigating strongly correlated photons \citep{Greentree2006, Hartmann2006, Angelakis2007, Carusotto2009, Hartmann2010}. Superconducting simulators \citep{Houck2012, Schmidt2012} use transmon qubits on every lattice site to induce photon repulsion \citep{Raftery2014, Fitzpatrick2017} using the Jaynes-Cummings nonlinearity, where parameter control increases in complexity with system size. A hardware-efficient approach would be to build a field-programmable simulator by using a single nonlinear circuit element to couple a blockade mode to an array of linear oscillators to allow site-selective parametric control over the magnitude of photon interactions.

\begin{figure*}[!t]
	\begin{center}
		\includegraphics[width=1.0\textwidth]{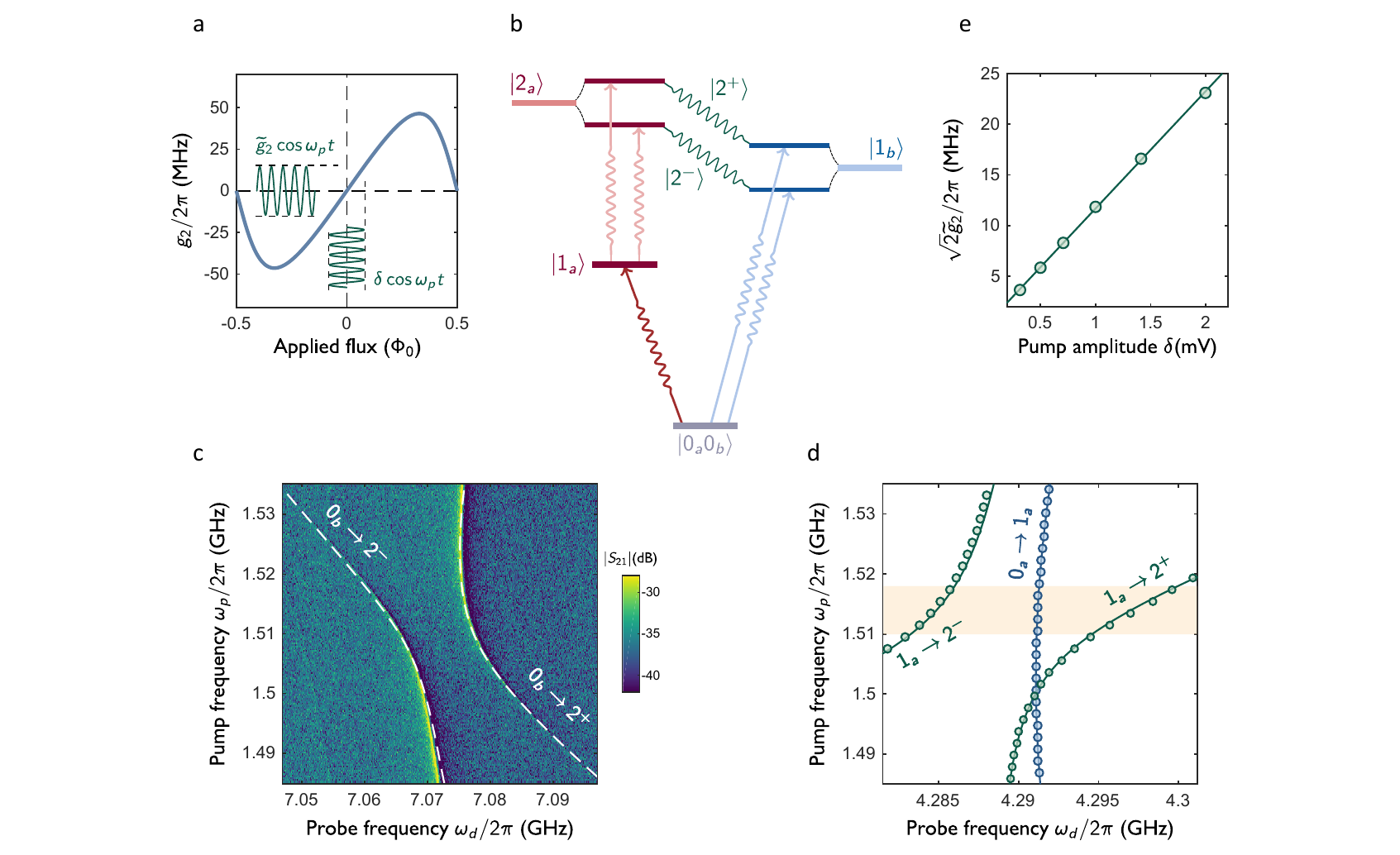}
	\end{center}
	\vspace{-0.6cm}
	\caption{\label{fig:TWM} \textbf{Dynamical three-wave interaction.} \textbf{a.} Flux dependence for the three-wave coupling term $g_2$ calculated using the fitted circuit parameters. Flux modulation at the sweet spot yields a modulated coupling amplitude. \textbf{b.} Energy level diagram for the coupled mode system outlining the stimulated nonlinearity arising from the dynamical Rabi splitting of the logical $|2_a\rangle$ state. The dynamical eigenstates $|2^\pm\rangle$ are probed by measuring their blockade component with a single photon probe (blue) or the logical component using a pump and probe (red). \textbf{c.} Measured transmission $|S_{21}|$ through the blockade resonator as a function of probe and flux pump frequency. The normal mode splitting was fitted (dashed line) in order to extract the dynamical three-wave coupling strength. \textbf{d.} Extracted resonance peaks in the pump-probe spectroscopy performed on the logical resonator. The avoided crossing was fitted (solid line) for confirming the magnitude of the oscillator anharmonicity $\sqrt{2}\tilde{g}_2/2\pi$. Shaded region: optimum flux pump frequency range. \textbf{e.} Measured stimulated three-wave coupling as a function of flux modulation amplitude, following a linear dependence (solid line).} 
\end{figure*}

This experiment is implemented using a circuit quantum electrodynamics (cQED) architecture \citep{Wallraff2004} as shown in Fig.~\ref{fig:Device}. The circuit consists of two superconducting microwave oscillators coupled inductively through a Josephson junction connected at the two voltage nodes. The resonators are composed of lumped element inductors and capacitors, and can be described quantum mechanically by their reduced node fluxes $\boldsymbol\varphi_{a,b}$. The closed loop composed of the resonator inductors and Josephson junction constitutes the circuit for a superconducting quantum interference device (rf-SQuID) which allows for the galvanic coupling between the resonator mode currents to be tuned with an external magnetic flux $\Phi_{\mathrm{ext}}$. Due to fluxoid quantization inside the loop, the coupler degree of freedom given by the phase difference across the junction can be written in terms of the cavity degrees of freedom, thereby reducing the complexity of the full circuit analysis (see Supplemental Information). A key ingredient in this experiment is the wave mixing capability of the Josephson junction captured in its inductive energy

\begin{align}\label{eqJJpotential}
\nonumber \boldsymbol{\mathcal{U}_{J}} &=-E_J \cos\left(\boldsymbol\varphi_a-\boldsymbol\varphi_b+\frac{2\pi}{\Phi_0}\Phi_{\mathrm{ext}}\right) \\
& = \xi_1 \widetilde{\boldsymbol\varphi}_{ab} + \xi_2 \widetilde{\boldsymbol\varphi}^2 _{ab}+  \xi_3 \widetilde{\boldsymbol\varphi}^3_{ab} +  \xi_4\widetilde{\boldsymbol\varphi}^4_{ab} +\mathcal{O}(\tilde{\boldsymbol\varphi}_{ab}^5)
\end{align}

\noindent where $E_J$ is the Josephson energy and $\Phi_0$ is the magnetic flux quantum. The phase difference between the cavity nodes is defined as $\boldsymbol\varphi_{ab} = \boldsymbol\varphi_{b} - \boldsymbol\varphi_{a}$. The junction nonlinearity becomes apparent in the second line of equation \eqref{eqJJpotential} obtained from Taylor expanding the potential about the minimum $\overline{\boldsymbol\varphi}_{\mathrm{ab}}$ and arriving at an effective potential for the phase deviation $\widetilde{\boldsymbol\varphi}_{\mathrm{ab}} = \boldsymbol\varphi_{\mathrm{ab}} -  \overline{\boldsymbol\varphi}_{\mathrm{ab}}$. The flux dependence has been absorbed in the expansion coefficients.

By itself, a Josephson junction has an even potential and below its plasma frequency can be modeled to second-order $\tilde{\boldsymbol{\varphi}}^2$ as a linear inductor. The fourth-order correction from the cosine potential $\tilde{\boldsymbol{\varphi}}^4$ gives a Kerr nonlinearity used extensively for engineering photon interactions \citep{Dykman2012}. Enforcing fluxoid quantization, in the rf-SQuID topology, generates an odd potential which ultimately yields the $\tilde{\boldsymbol{\varphi}}^3$ nonlinearity necessary for engineering the three-wave interaction. Other Josephson circuits with cubic nonlinearity have been implemented for applications in non-degenerate parametric amplification \citep{Frattini2017}, driving forbidden transitions in an artificial $\Lambda$-system \citep{Vool2018}, and quantum simulation of the ultrastrong coupling regime \citep{Markovic2018}.

The circuit is described by the effective Hamiltonian obtained from promoting the node flux and conjugate charge  variables to quantum operators 

\begin{align}\label{eqHamiltonian}
\nonumber \boldsymbol{\mathcal{H}}/\hbar=\,\,\,  &\omega_a \op{a}^\dagger \op{a} +\omega_b  \op{b}^\dagger \op{b} + g (\op{a}^\dagger \op{b} + \op{a} \op{b}^\dagger) + g_2 (\op{a}^{\dagger 2} \op{b} + \op{a}^2 \op{b}^\dagger)\\
&+\chi_{ab} \op{a}^\dagger \op{a} \op{b}^\dagger \op{b} + \frac{\chi_{aa}}{2} \op{a}^{\dagger 2} \op{a}^2 + \frac{\chi_{bb}}{2} \op{b}^{\dagger 2} \op{b}^2 .
\end{align}

\noindent The operators $\op{a}^\dagger (\op{a})$ and $\op{b}^\dagger (\op{b})$ create (annihilate) photons in the logical (blockade) eigenmodes, where $\omega_a$ and $\omega_b$ are the mode frequencies. The linear Josephson inductance contributes to the total inductance of each resonator and provides a linear coupling between the modes at a rate $g$ smaller than their detuning, effectively adding dispersive corrections to the mode frequencies. The amplitude $g_2$ corresponds to the rate of exchanging pairs of logical photons with single blockade photons. The second line of equation (\ref{eqHamiltonian}) takes into account the Kerr nonlinearity that both modes inherit from the junction. For brevity, all fast rotating terms, apart from the three-wave term, have been removed from the Hamiltonian, however, the full Rabi model is used in theoretical analysis and modeling (see Supplemental Information). Since all coefficients in equation (\ref{eqHamiltonian}) depend on the applied magnetic flux, our circuit parameters are chosen ($L_J>L_{a,b}$) to ensure the mode frequencies have a weak flux dependence (Fig.~\ref{fig:Device}D). In order to minimize oscillator dephasing, we bias the rf-SQuID at zero magnetic flux. At this operating point the resonator frequencies are $\omega_a/2\pi=4.284$ GHz and $\omega_b/2\pi=7.073$ GHz, and their self-Kerr nonlinearities are $\chi_{aa}/2\pi = 3.0$ MHz and $\chi_{bb}/2\pi = 12.5$ MHz, respectively. The cross-Kerr nonlinearity $\chi_{ab}/2\pi = 10$ MHz will be used as a measurement tool for characterizing the logical mode state by probing transmission through the blockade resonator. Unlike the ubiquitous transmon circuit \citep{Koch2007}, the logical oscillator does not rely on the self-Kerr nonlinearity $\chi_{aa}$ since it is two orders of magnitude smaller than in the transmon case. We will rely on the lower-order cubic interaction $g_2$ for generating anharmonicity. Since the three-wave coupling rate $g_2$ is exactly zero at zero flux bias (Fig.~\ref{fig:TWM}A), we cannot rely on the static interaction to activate anharmonicity.

Nevertheless, this compromise between coherence and addressability can be overcome by parametric driving. This is accomplished by applying a sinusoidal modulation $\delta \cos \omega_p t$ to the rf-SQuID flux bias which effectively turns all terms in equation (\ref{eqHamiltonian}) into time-dependent periodic functions. To first order in flux, the three-wave coupling amplitude can be approximated as $g_2 \approx\tilde{g}_2(\delta) \cos\omega_pt + \mathcal{O}(\delta^2)$. Although the static coupling is zero, we can take advantage of the linear flux dispersion around the sweet spot and induce a non-zero dynamical amplitude $\tilde{g}_2 (\delta)= \delta \partial g_2/\partial \Phi$ that depends on the slope for small modulation amplitudes $\delta/\Phi_0 \ll 1$. For larger amplitudes $\delta/\Phi_0 \sim 0.1$, using the Jacobi-Anger expansion (see Supplemental Information), the coupling becomes $\tilde{g}_2(\delta) = 2J_1(\delta) g_2$, where $J_1$ is the first Bessel function. By appropriately choosing the pump frequency to approximately match $\omega_p = 2\omega_a - \omega_b$, the three-wave interaction in the frame rotating at the oscillator frequencies $\tilde{g}_2(\delta)\left(\op{a}^{\dagger 2}\op{b} + \mathrm{h.c.}\right)$ becomes effectively resonant. The flux modulation therefore stimulates a frequency conversion process which exchanges a pair of logical photons with a single blockade photon. This coherent interaction leads to new eigenstates defined as the symmetric and antisymmetric superposition of the two oscillator excitations $|2^\pm\rangle = (|2_a0_b\rangle \pm |0_a1_b\rangle) /\sqrt{2}$ (Fig.~\ref{fig:TWM}B). Since the oscillator frequencies have a non-linear flux dependence near the sweet spot, flux modulation will shift the modes by $\Delta\omega = -{\delta^2}/{4}({\partial^2 \omega}/{\partial \Phi^2})$. These second-order corrections to the measured resonator frequencies have also been observed in other experiments\citep{Naik2017, McKay2016, Reagor2018, Bajjani2011} where flux modulation was used in an analogous fashion, and need to be taken into consideration when calibrating the parameters of the flux drive.

\begin{figure*}[!t]
	\begin{center}
		\includegraphics[width=1.0\textwidth]{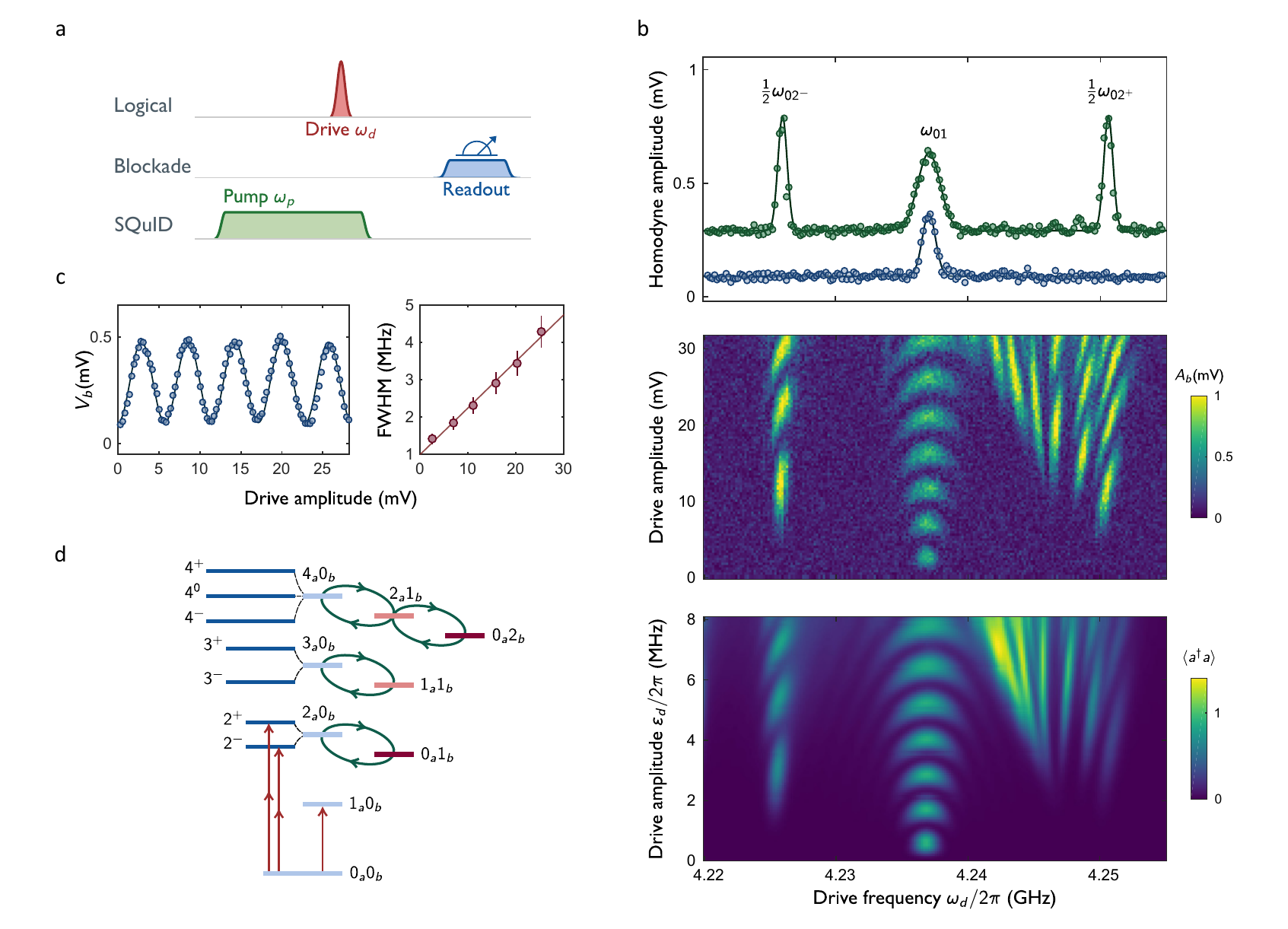}
	\end{center}
	\vspace{-0.6cm}
	\caption{\label{fig:Rabi} \textbf{Time domain spectroscopy.} \textbf{a.} Experimental pulse sequence for probing the anharmonic logical mode. The SQuID coupler is driven with a $10\mu$s pulse with a microwave carrier frequency $\omega_p/2\pi = 1.557$ GHz and amplitude $\delta\simeq 2$ mV to reach a three-wave coupling strength $\sqrt{2}\tilde{g}_2/2\pi \simeq 25$ MHz. The logical cavity is driven with a narrow bandwidth Gaussian pulse ($4\sigma = 1\mu$s) in order to resolve all spectroscopic transitions. The blockade cavity is weakly probed at its resonance frequency when the logical mode is in vacuum, $\omega_b/2\pi =7.037$ GHz. \textbf{b.} Probing of the logical cavity spectrum using the transmitted blockade homodyne voltage $A_b$. Top: spectroscopic features for two different drive amplitudes, $2.53$ mV (blue) and $11.07$ mV (green), showing a power broadened single-photon resonance and the emergence of a two-photon resonance. Middle: measured readout signal versus logical drive frequency and amplitude. Bottom: Theoretical prediction based on master equation simulation. \textbf{c.} Left: Rabi oscillations in the homodyne voltage probed at the single-photon peak $\omega_{01}/2\pi$. Right: fitted peak full width at half maximum (FWHM) showing linear dependence with drive amplitude (solid line) as predicted by Bloch equations. \textbf{d.} Energy level diagram for the two cavity eigenstates with parametric three-wave coupling (green arrows) where higher-lying eigenstates are probed as multi-photon transitions (red arrows).} 
\end{figure*}

This dynamically induced nonlinearity is very sensitive to the selected pump frequency. The simplest calibration experiment is done in a continuous wave form by probing the transmission spectrum of the blockade resonator with a weak coherent tone and applying a flux pump tone at a fixed modulation amplitude.  As we sweep both probe and flux pump frequencies, we get a well-resolved normal-mode splitting centered at $\omega_p = 2\omega_a - \omega_b \approx 2\pi\times 1.510$ GHz (Fig.~\ref{fig:TWM}C). This is clear evidence that we are operating in the strong conversion limit $\sqrt{2}g_2 > \kappa_a,\kappa_b$,  where the three-wave coupling is stronger than the logical (blockade) dissipation rate $\kappa_a (\kappa_b)$. This experiment involved probing the overlap of a single blockade photon with the dynamically hybridized states $|2^\pm\rangle$. Additionally, we can also probe the logical two-photon component by climbing the energy ladder using a pulsed pump-probe scheme while continuously applying the same flux pump tone. The first pulse populates the $|1_a\rangle$ state by resonantly driving the logical mode and, subsequently, the transmission spectrum of the logical resonator is probed using a second pulse. A similar avoided crossing is observed (Fig.~\ref{fig:TWM}D) when population is transferred to the logical two-photon manifold, specifically when the probe frequency matches the $|1_a\rangle \rightarrow |2^\pm\rangle$ transitions. Both measurement approaches yield the same value for the dynamically activated three-wave coupling $\tilde{g}_2$; however, the latter provides spectroscopic evidence that for an optimal window in flux pump frequencies we can drive the logical mode as a qubit since the energies required to leak into higher-lying eigenstates are significantly detuned from the single-photon transition. This calibration scheme is repeated for pump amplitudes up to $\delta/\Phi_0 \sim 0.2$, yielding nonlinearities as large as $\sqrt{2}\tilde{g}_2/2\pi \sim 25$ MHz (Fig.~\ref{fig:TWM}E), where the dynamical coupling strength increases linearly with the pump amplitude, as expected. The pump amplitude was not further increased to avoid generating higher harmonics of the pump frequency as we deviate from the linear flux dispersion of $g_2(\Phi_\mathrm{ext})$.

To demonstrate that the multi-level logical oscillator can be reduced to a qubit, the following experimental sequence is employed (Fig.~\ref{fig:Rabi}A). First, the flux is modulated at the optimal pump frequency $\omega_p$ and amplitude $\delta$ which activates the anharmonicity. Subsequently, the logical resonator is excited with a drive pulse of varying amplitude and frequency. Finally, we measured the homodyne amplitude of a probe tone transmitted through the blockade cavity.
Due to the cross-Kerr nonlinearity $\chi_{ab} \op{a}^\dagger \op{a} \op{b}^\dagger \op{b}$  resolved at the single photon level ($\chi_{ab} > \kappa_a,\kappa_b$), the blockade resonance acquires a frequency shift dependent on the number of photons in the logical mode \citep{Holland2015}, very similar to the state dependent shift between a qubit and a cavity in the strong dispersive regime \citep{Schuster2007}. This allows indirect measurement of the logical mode population by weakly probing the blockade resonator transmission. Since the three-wave hybridization would further complicate the measurement, flux modulation is turned off before readout. The outcome of measuring the homodyne amplitude of the blockade tone as a function of drive amplitude and frequency is shown in Fig.~\ref{fig:Rabi}B. At low drive amplitudes we observe a resonance peak when the drive frequency is tuned near the single-photon transition frequency $\omega_{01}/2\pi$ and by further increasing the drive amplitude we observe clear oscillations in the peak height and a strong amplitude dependence of the line width (Fig.~\ref{fig:Rabi}C). This is consistent with observing Rabi oscillations and power broadening for a resonantly driven spin 1/2 described by the Bloch equations. The effective spin in this case corresponds to the energetically isolated single-photon manifold of the logical mode.

Higher eigenstates of the coupled system become accessible with increasing drive amplitude via multi-photon transitions \citep{Bishop2009}, since direct transitions from vacuum are suppressed by parity selection rules. These eigenstates are significantly perturbed under flux modulation as the dynamical three-wave interaction mixes states in different photon number manifolds (Fig.~\ref{fig:Rabi}D). Transitions to the $|2^\pm\rangle$ states occur as two photon oscillations with resonance frequencies $\frac{1}{2}\omega_{02^\pm}$ symmetrically detuned from the single-photon transition frequency $\omega_{01}$. The spectroscopic features red-detuned from the two-photon resonance $\frac{1}{2}\omega_{02^+}$ are multi-photon oscillations.

We model the dynamics of the Hamiltonian in equation (\ref{eqHamiltonian}) when it is subjected to a coherent drive on the logical cavity $\mathcal{H}_d = \varepsilon_d(t)\op{a}^\dagger + \mathrm{h.c.}$ and a time dependent flux drive. Moving to a rotating frame that combines both the coherent drive frequency $\omega_d/2\pi$ and flux pump frequency $\omega_p/2\pi$, the Hamiltonian becomes $\mathcal{H} = \Delta_a \op{a}^\dagger \op{a} + \Delta_b \op{b}^\dagger \op{b}  + \tilde{g}_2(t)(\op{a}^{\dagger 2}\op{b} + \mathrm{h.c.}) + \mathcal{H}_\mathrm{Kerr}  + \mathcal{H}_\mathrm{d}$ where $\Delta_a = \omega^\prime_a - \omega_d$ and $\Delta_b = \omega^\prime_b - 2\omega_d + \omega_p$. The non-resonant linear coupling terms are eliminated using a Schrieffer-Wolff transformation and $\omega^\prime_{a,b}$ represent the dressed oscillator frequencies. Additionally, this model accounts for mode frequency corrections from the second harmonic of the flux tone. In order to study the dynamics of the reduced density matrix $\rho$, losses are incorporated by employing the quantum master equation written in the standard Lindblad form

\begin{align}\label{eqLindblad}
\dot{\rho} = -\frac{i}{\hbar}[\boldsymbol{\mathcal{H}},\rho] + \kappa_a\mathcal{D}[\op{a}]\rho + \kappa_b\mathcal{D}[\op{b}]\rho
\end{align}

\noindent where we have the usual definition for the Lindblad damping superoperator $\mathcal{D}[\mathbf{L}]\rho = \mathbf{L}\rho\mathbf{L}^\dagger - \frac{1}{2}\mathbf{L}^\dagger\mathbf{L}\rho - \frac{1}{2}\rho\mathbf{L}^\dagger\mathbf{L}$ used for modeling the photon loss in the two resonators. The time-dependent envelopes for the coherent drive and three-wave interaction amplitudes have been modeled to replicate the experimentally implemented pulse sequence. At the end of every pulse instance  we calculate the expectation value of the logical cavity population $\mathbf{Tr}(\rho \op{a}^\dagger \op{a})$ as a function of the drive amplitude and frequency. As shown in Fig.~\ref{fig:Rabi}B, this theoretical framework clearly captures the experimentally measured Rabi oscillations and power broadening at the single-photon resonance. The numerical simulations also confirm that the leftmost and rightmost resonances belong to two-photon transitions to the states $\vert 2^-\rangle$ and $\vert 2^+\rangle$ respectively, and show that the spurious resonances located between $\omega_{01}$ and $\frac{1}{2}\omega_{02^+}$ in the drive frequency range correspond to complex multi-photon transitions  involving the excitation of the states $\vert 2^+\rangle, \vert 3^+\rangle, \vert 4^+\rangle$ and $\vert 5^+\rangle$. We expect that the discrepancies in the multi-photon oscillation periods arise from distorted pulse waveforms in the experimental setup, which are not fully taken into consideration in our theory.

\begin{figure}[!t]
	\begin{center}
		\includegraphics[width=0.5\textwidth]{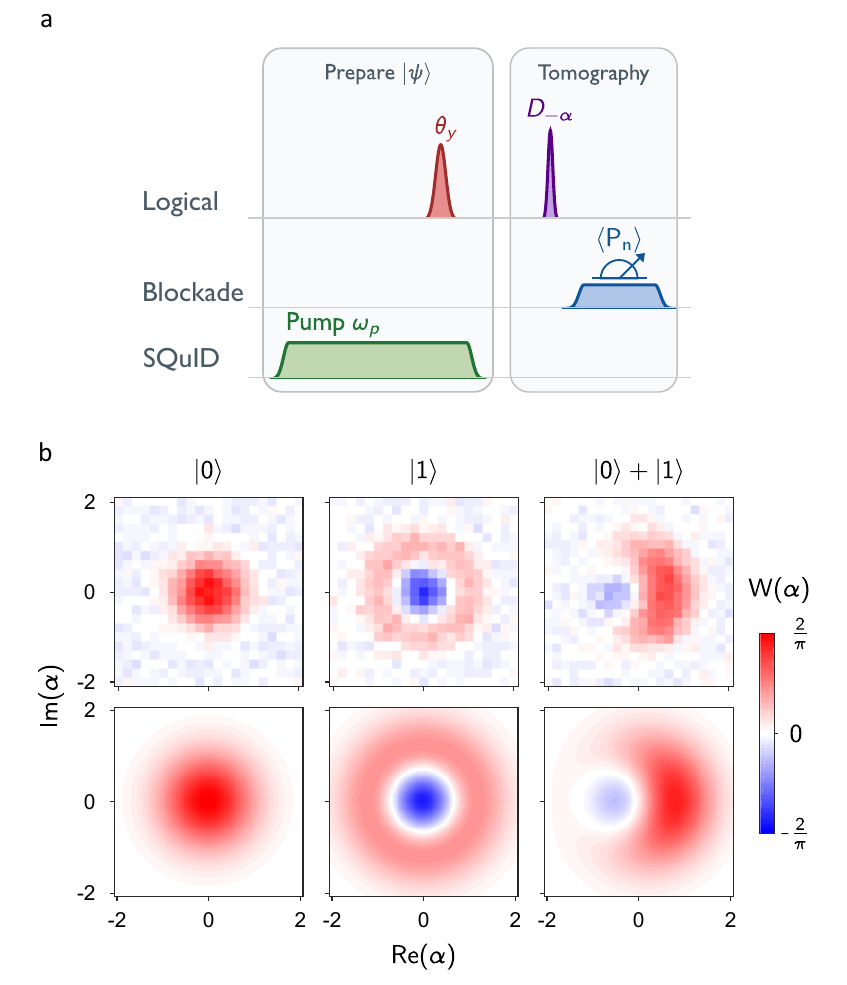}
	\end{center}
	\vspace{-0.5cm}
	\caption{\label{fig:WignerTomo}\textbf{Wigner tomography.} \textbf{a.} Experimental pulse sequence for characterizing the logical resonator state. The preparation scheme is the same as in Fig.~\ref{fig:Rabi}A. The logical cavity is driven with a $4\sigma = 100$ns Gaussian pulse. The tomography sequence consists of a 10 ns resonant pulse, which displaces the prepared state by $-\alpha$, and a measurement pulse through the blockade cavity conditioned on having $n$ photons in the logical cavity, probing the displaced state in Fock basis  $p_n = \langle n|\mathbf{D}_{-\alpha}|\psi\rangle$. \textbf{b.} Wigner functions for the prepared logical states $|0\rangle, |1\rangle, |+\rangle$. Top: Experimental results. Bottom: Theoretical prediction based on master equation simulation.} 
\end{figure}

Following the Rabi experiment, we can prepare arbitrary states $|\psi\rangle = \alpha |0\rangle + \beta|1\rangle$ in the single-photon manifold of the logical oscillator. Specifically, we calibrate the amplitude of the drive pulse to perform $(\pi)_y$ and $(\pi/2)_y$ rotations in the logical subspace. Using the protocol described in Fig.~\ref{fig:WignerTomo}A, similar to the approach in refs  \citep{Hofheinz2009,Kirchmair2013}, we perform tomography on the cavity state by measuring the Wigner function

\begin{align}\label{eqQn}
\mathit{W}(\alpha) = \frac{2}{\pi}\langle \psi\lvert\mathbf{D}_\alpha \mathbf{P} \mathbf{D}^\dagger_\alpha \rvert \psi\rangle
\end{align}

\begin{figure}[!t]
	\begin{center}
		\includegraphics[width=0.5\textwidth]{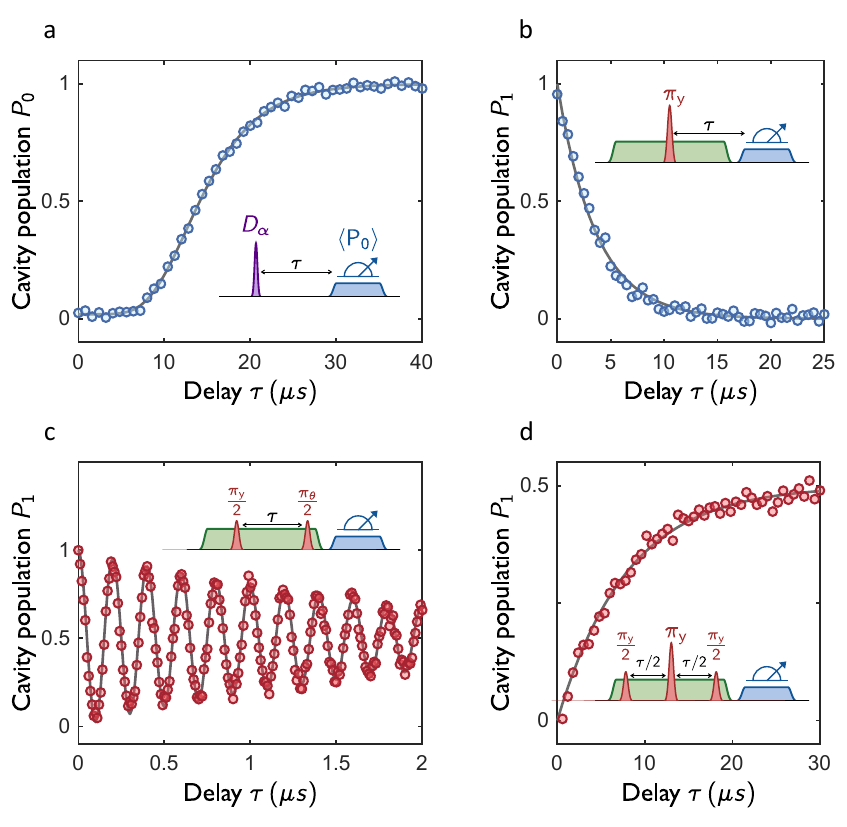}
	\end{center}
	\vspace{-0.5cm}
	\caption{\label{fig:Coherence}\textbf{Oscillator coherence.} \textbf{a.} Measuring the cavity decay rate $\kappa_a$ by probing the vacuum population after displacing the cavity. \textbf{b.} Single-photon energy decay. $|1_a\rangle$ state is prepared using a calibrated $\pi_y$ pulse. \textbf{c.} Ramsey experiment to measure phase coherence by applying two resonant $(\pi/2)$ pulses with a delay $\tau$ in between. The fringe frequency $\omega_\mathrm{fr}$ is set artificially by varying the azimuth of the second pulse by $\Delta \theta = \omega_\mathrm{fr}\tau$. \textbf{d.} Spin-echo decay measurement. All insets reveal the pulse sequences used for each measurement.} 
\end{figure}

\noindent defined in this form as the expectation value of the parity operator $\mathbf{P} = e^{i\pi\op{a}^\dagger\op{a}}$ evaluated over the complex plane spanned by the displacement operator $\mathbf{D}_\alpha = e^{\alpha \op{a}^\dagger- \alpha^\ast \op{a}}$. The tomography sequence starts with resonantly driving the logical mode with a short Gaussian pulse which can be approximated as a coherent displacement for a weakly anharmonic ($\chi_{a} = 3$ MHz) logical mode, provided that the spectral bandwidth is large enough to resonantly drive many nearest level transitions \citep{Shalibo2013}. The size and phase of the complex displacement $\alpha$ is controlled by the amplitude and phase of the displacement pulse. After displacing the prepared state by $-\alpha$ we measure the expectation value of the parity operator $\langle \mathbf{P} \rangle = \pi \sum_{n}(-1)^n P_n$ expressed in terms of the photon number occupation probabilities $P_n$. Using the photon number dependent frequency shift of the blockade resonance, the probability of having $n$ photons in the logical mode is proportional to the transmitted homodyne signal through the blockade cavity probed at $\omega_b - n\chi_{ab}$.  The displacement pulse amplitude is calibrated in units of $|\alpha|$ by fitting the number-selective readout voltage to a Poisson distribution which also normalizes the measured homodyne signal to occupation probabilities for photon numbers up to $n = 3$. The experimentally measured Wigner functions are shown in Fig.~\ref{fig:WignerTomo}B for the prepared cavity states $|0\rangle$, $|1\rangle$ and $|+\rangle  = (|0\rangle + |1\rangle)/\sqrt{2}$ respectively. Performing least-squares regression on the measured Wigner functions we can reconstruct the density matrix $\rho$ for the logical mode (see Supplementary Information). This density matrix is then used to calculate the state fidelity $\mathcal{F} = \langle\psi|\rho|\psi\rangle$ for all experimentally prepared states: $\mathcal{F}_{|0\rangle} = 0.94 \pm 0.009$, $\mathcal{F}_{|1\rangle} = 0.90 \pm 0.002$ and $\mathcal{F}_{|+\rangle} = 0.831 \pm 0.048$. The fidelity for preparing the ground state is limited by thermal occupation.

The oscillator coherence is bounded by the classical energy decay rate $\kappa_a$ at the single photon level. This was measured in the absence of flux modulation. We prepare the logical mode in a coherent state with $\alpha_0 \simeq 3$ using a calibrated displacement pulse and after a delay time $\tau$, we measure the population of the vacuum state $P_0(\tau)$ \citep{Reagor2016} by probing the number-selective transmission of the blockade resonator. The initial state decays as $\alpha(t) = \alpha_0 \exp(-\kappa t)$ and following Poisson statistics, the probability of measuring vacuum is $P_0(\tau) = \exp(-|\alpha(\tau)|^2)$. Fitting the double exponential decay to the readout signal (Fig.~\ref{fig:Coherence}A) we extract $\kappa_a/2\pi = 35.2\pm 0.3\,\mathrm{kHz}$ corresponding to a lifetime $T_\kappa = 1/\kappa = 4.52\pm0.04\,\mathrm{\mu s}$. 

The relaxation time $T_1$ of a resonator can be characterized by preparing the $|1\rangle$ state and measuring the probability of finding the oscillator in vacuum as a function of delay. We find a $T_1 = 4.0\pm 0.22\,\mathrm{\mu s}$ (Fig.~\ref{fig:Coherence}B), which is in close agreement with the measured classical decay rate of a coherent state. Based on this result we speculate that the measured fidelity $\mathcal{F}_{|1\rangle}$ of preparing a single photon is limited by relaxation. Similarly, using a Ramsey-type measurement,  we can prepare the $(|0\rangle+|1\rangle)/\sqrt{2}$ state and measure the decay of phase coherence as well as calibrate the $|0\rangle \rightarrow |1\rangle$ transition frequency. The measured decay rate $T_{2\ast} = 2.1\pm0.17\,\mathrm{\mu s}$ (Fig.~\ref{fig:Coherence}C) effectively translates to a dephasing time $T_\phi = 2.91\pm 0.32\,\mathrm{\mu s}$. We believe this dephasing rate to be limited by low-frequency flux noise which was filtered out in a single-echo measurement to get a relaxation-limited decoherence time $T_{2\mathrm{e}} = 7.95\pm0.82\,\mathrm{\mu s} \approx 2T_1$ (Fig.~\ref{fig:Coherence}D). We calculate relaxation and dephasing rates from various noise sources affecting our circuit (see Supplemental Information). Based on estimates for the loss tangent at the material interface for planar resonators and transmon qubits \citep{Wenner2011, Wang2015}, we find $T_1$ to be limited by dielectric loss present in the resonator capacitance to ground. This loss can be reduced by three orders of magnitude using three-dimensional cavities which are less sensitive to dielectric and conductor loss \citep{Reagor2013}. The dephasing rate was found to be limited by the critical current noise in the coupler junction.

In conclusion, we have demonstrated the capability of tailoring the Hilbert space of a superconducting oscillator by dynamically activating a three-wave interaction with an ancillary mode. Owing to the large three-wave coupling, this platform can be advantageous to dissipative stabilization schemes\citep{Leghtas2015, Touzard2018}, which rely on two-photon loss for confining a state in a manifold protected against dephasing errors\citep{Mirrahimi2014}. For our purpose, this stimulated nonlinearity allows the single-photon manifold to be selectively addressed as an effective two-level system with photon lifetimes limited by the resonator's intrinsic quality factor. Further optimizing the circuit parameters, it is possible to engineer a cavity anharmonicity similar to that of the transmon. Despite the modest coherence of lithographically defined resonators, this proof of principle can be extended to three-dimensional microwave cavities for achieving qubits with milisecond \citep{Reagor2016,Reagor2013}, up to second \citep{Romanenko2018}, coherence times.

\subsection*{Acknowledgments}
The authors thank M. Mirrahimi and J. A. Aumentado for valuable discussions, G. Zhang and P. Mundada for technical contributions, and MIT Lincoln Labs for providing a travelling wave parametric amplifier for this experiment. This work was supported by the Army Research Office through grant no. W911NF-15-1-0421.

\subsection*{Author contributions}
A.V. designed the device, performed the experiments, and analyzed the data. Z.H., P.G. and J.K. provided theoretical support. A.A.H supervised the whole experiment. All authors contributed to the preparation of this manuscript.

\end{document}